\documentclass[12pt,a4paper]{article}

\usepackage[utf8]{inputenc}
\usepackage[english]{babel}
\usepackage{amsmath}
\usepackage{amsfonts}
\usepackage{amssymb}

\usepackage[comma,numbers,sort&compress]{natbib}
\usepackage{graphicx}
\usepackage[left=2cm,right=2cm,top=2cm,bottom=2cm]{geometry}

\author{Michael Florian Wondrak$^{a,b,}$\footnote{wondrak@fias.uni-frankfurt.de}\\
\small $^a$ Frankfurt Institute for Advanced Studies (FIAS)\\
\small Ruth-Moufang-Str.~1, 60438 Frankfurt am Main, Germany\\[1ex]
\small $^b$ Institut f\"{u}r Theoretische Physik, Johann Wolfgang Goethe-Universit\"{a}t Frankfurt\\
\small Max-von-Laue-Str.~1, 60438 Frankfurt am Main, Germany}
\date{}
\title{The Cosmological Constant and Its Problems:\\A Review of Gravitational Aether}

\begin{document}
\maketitle

\begin{abstract}
\noindent In this essay we offer a comprehensible overview of the gravitational aether scenario. This is a possible extension of Einstein's theory of relativity to the quantum regime via an effective approach. Quantization of gravity usually faces several issues including an unexpected high vacuum energy density caused by quantum fluctuations. The model presented in this paper offers a solution to the so-called cosmological constant problems. 

As its name suggests, the gravitational aether introduces preferred reference frames, while it remains compatible with the general theory of relativity. As a rare feature among quantum gravity inspired theories, it can predict measurable astronomical and cosmological effects. Observational data disfavor the gravitational aether scenario at $2.6\text{--}5\,\sigma$. This experimental feedback gives rise to possible refinements of the theory.
\end{abstract}

\section{Introduction}
Probably everyone of us has experienced the benefits of satellite-based navigation systems, e.g.~in route guidance systems for cars. The incredible accuracy to determine one's position is only enabled by consideration of the effects of general relativity. In everyday life this is an omnipresent evidence for the performance of Einstein's theory. Nevertheless, this theory is classical and could not yet be merged with quantum theory. This is important since quantum fluctuations may determine the expansion behavior of the universe. The gravitational aether is a phenomenological concept to effectively address this issue as an ubiquitous field interacting with ordinary matter. In this essay we want to examine the gravitational aether scenario as a testable theory inspired by quantum gravity.

We begin with a short recap of the standard model of cosmology in section \ref{sec:CCP_EinsteinAndTheCosmologicalConstant} before we focus on the cosmological constant and its problems in section \ref{sec:CCP_CosmologicalConstantProblems}. Section \ref{sec:CCP_GravitationalAetherScenario} introduces the gravitational aether concept and its repercussions on the cosmological constant problems. Section \ref{sec:CCP_ExperimentalTests} is devoted to testable predictions of this theory, comparison with observational data, and possible improvements. We draw our conclusions in section \ref{sec:CCP_SummaryAndOutlook}.

\section{Einstein and the cosmological constant}
\label{sec:CCP_EinsteinAndTheCosmologicalConstant}
The revolutionary idea behind the general theory of relativity (GR) is to encode gravity in the geometry of the universe, more precisely, in its metric tensor $g_{\mu\nu}$ which serves as a local ruler telling us about deformations of space and time. The geometry is determined by the distribution of matter, or more generally, by the distribution of energy but specifies the evolution of matter in return. This interplay manifests itself in the so-called Einstein field equations which date back to 1915,
\begin{equation}
G_{\mu\nu} +\Lambda_\text{b}\,g_{\mu\nu} = \frac{8\pi\,G_\text{N}}{c^4}\,T_{\mu\nu}.
\label{eq:EFE_wGwoL}
\end{equation}
The Einstein tensor $G_{\mu\nu}$ is a function of $g_{\mu\nu}$ and represents the geometric part while the energy-momentum tensor $T_{\mu\nu}$ encodes information about the energy distribution. $\mu$ and $\nu$ denote the respective tensor components. $\Lambda_\text{b}$ is the so-called bare cosmological constant, $G_\text{N}$ denotes the gravitational constant, and $c$ stands for the speed of light which we will set to unity, $c \equiv 1$, following the high-energy physics convention. 

We can apply the Einstein field equations to describe the evolution of the universe as a whole. This is subject to cosmology. Already in 1924 Alexander Friedmann derived evolution equations of the universe's scale factor which allow a variety of different scenarios: forever expanding, static, or collapsing universes with open (negatively curved), flat or closed (positively curved) geometries. Einstein believed in a static spacetime and introduced the cosmological constant to compensate the attractive forces of matter. Investigating the relation between redshifts and distances of galaxies, Edwin Hubble concluded in 1929 that our universe is expanding. This contradiction to his original assumption caused Einstein to call his insertion the biggest blunder of his life, as is reported by George Gamow \citep{Gamow1970}.

By now, new sources of cosmological information have been tapped, most prominently the cosmic microwave background radiation (CMB), baryon acoustic oscillations (BAO), and standard candles, i.e. cosmic objects or events like supernova Ia explosions (SNe) whose distance from us can be measured very precisely. In the future, gravitational wave (GW) observations could extend this spectrum. In 1998 SNe observations indicated that the expansion of the universe nowadays is even accelerated \citep{Perlmutter1998,Riess1998}. 

According to the standard model of cosmology, the $\Lambda$CDM model, we suppose that the universe has formed in a hot big bang about 13.8 billion years ago. Shortly after its formation it underwent a phase of very fast expansion (inflation phase) which wiped out inhomogeneities leaving us with a nearly flat spacetime. The temperature of the universe decreased as it expanded. At first, radiation was the dominant constituent, followed by non-relativistic matter and finally by dark energy which is responsible for the accelerated expansion. The name $\Lambda$CDM is composed of $\Lambda$ which stands for the cosmological constant as a specific type of dark energy and CDM denoting cold dark matter. Together with the ordinary so-called baryonic matter the latter builds up non-relativistic matter.

In order to proof that the cosmological constant $\Lambda_\text{b}$ can lead to an accelerated expansion, we have a look at the second Friedmann equation,
\begin{equation}
\frac{\ddot{a}}{a}
= -\frac{4\pi\,G_\text{N}}{3}\,\sum_j\left(\rho_j +3p_j\right) 
  +\frac{\Lambda_\text{b}}{3}.
\end{equation}
Here the acceleration $\ddot{a}$ of the universe's scale factor $a$ is determined by the energy densities $\rho_j$ and pressure contributions $p_j$ of the different matter components $j$ present. The implication of a positive $\Lambda_\text{b}$ is to support an accelerated expansion, $\ddot{a}>0$, and finally to cause an exponential growth. (For the experts: The matter components are modeled as perfect fluids which differ in their equations of state $w=p/\rho$: for radiation $w=1/3$ and for non-relativistic matter $w=0$.)

\section{Cosmological constant problems}
\label{sec:CCP_CosmologicalConstantProblems}
General relativity is a classical theory in the sense that it is not quantized and so the Einstein field equations perform primely for classical matter fields. But we know well that matter fields are of quantized nature which implies that there is a persisting non-vanishing energy density even in the vacuum state, i.e.~if no particle is present. This vacuum energy density $\rho_\text{vac}$ has the same value at every instance of spacetime and its energy-momentum tensor is of perfect-fluid type \citep{Zeldovich1967,Sakharov1967}
\begin{equation}
T_{\text{vac},\,\mu\nu}
\equiv -\rho_\text{vac}\,g_{\mu\nu}.
\label{eq:EpTVac}
\end{equation}
In the spirit of GR, $T_{\text{vac},\,\mu\nu}$ is supposed to contribute to the total energy-momentum tensor $T_{\mu\nu}$ on the right-hand side of the Einstein field equations \eqref{eq:EFE_wGwoL} and thus to have an impact on spacetime geometry. Since it has the same structure as the term for the cosmological constant $\Lambda_\text{b}$ in the Einstein field equations, both lead to the same phenomena. They can be merged to yield the effective cosmological constant
\begin{equation}
\Lambda_\text{eff}
\equiv \Lambda_\text{b} +8\pi\,G_\text{N}\,\rho_\text{vac}.
\end{equation}

\begin{figure}[hbt]
\centering
\includegraphics[width=0.85\linewidth]{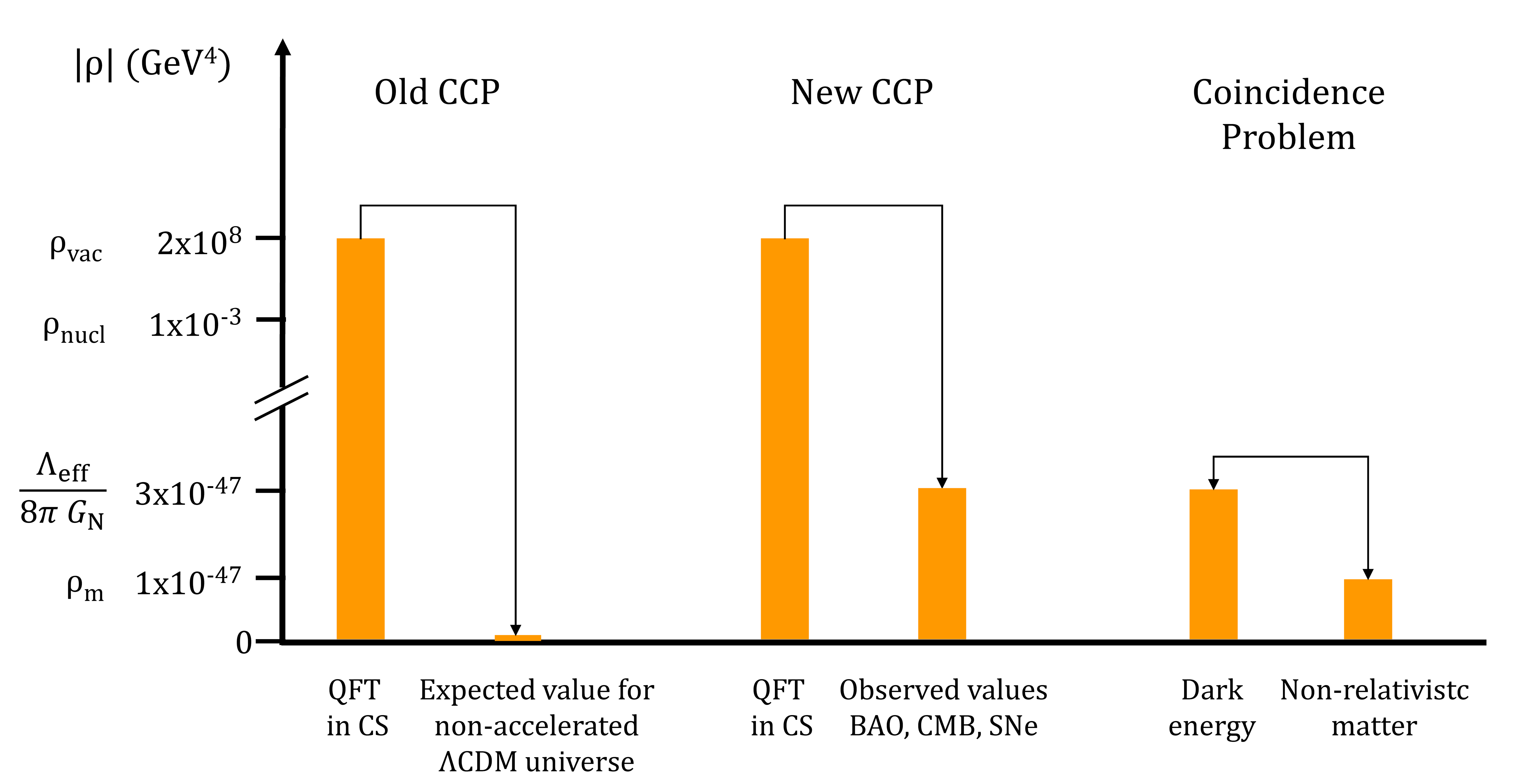}
\caption{Illustration of the cosmological constant problems (CCP).}
\label{fig:CCP}
\end{figure}

In principle, for an accurate determination of $\rho_\text{vac}$ one should take into account the fundamental nature of spacetime which is expected to be quantized -- thus one would need a theory of quantum gravity. Since such a theory is not yet available, we can employ semiclassical gravity. We assume a classical curved spacetime within which we define quantum fields. Those in return influence the spacetime geometry according to the Einstein field equations. For details we refer the willing reader to \citep{BirrellD1994,Martin2012}. As theoretical result we obtain
\begin{equation}
\rho_\text{vac}
= \sum_i {\left(-1\right)}^{2s_i}\,n_i\,\frac{m_i^4}{64\pi^2}\,
  \ln\!\left(\frac{m_i^2}{\mu^2}\right)
\end{equation}
up to possible contributions from phase transitions in the early universe. Here the sum runs over all fundamental quantum fields in the standard model of particle physics. They contribute according to their mass $m_i$, spin $s_i$, and number $n_i$ of degrees of freedom. 
$\mu$ is the renormalisation energy scale. Following \citep{KoksmaP2011}, $\mu$ can be related to the photons from SNe observations which are used to experimentally determine the cosmological constant. With $\mu\approx 3 \times {10}^{-25} \, \text{GeV}$ we find
\begin{equation}
\left| \rho_\text{vac} \right|
\approx 2 \times {10}^{8} \, \text{GeV}^4
\approx 5 \times {10}^{28} \, \frac{\text{kg}}{\text{m}^3}
\approx 2 \times {10}^{11} \, \rho_\text{nucl}
\end{equation}
The absolute value of the vacuum energy density is thus expected to be 11 orders of magnitude higher than the density $\rho_\text{nucl}$ of atomic nuclei and, roughly, neutron stars. This is remarkable because the energy density of neutron stars is believed to be the largest stable one before a collapse to a black hole. This obvious strong conflict with measurements is referred to as the so-called \textit{old cosmological constant problem} \citep{Weinberg1989}. This is illustrated in Fig.~\ref{fig:CCP}. In the past, physicists assumed that $\rho_\text{vac}$ would be exactly compensated by $\Lambda_\text{b}$ due to an unknown symmetry which would lead to a vanishing effective cosmological constant $\Lambda_\text{eff}$.

As mentioned above, modern cosmological observations support the idea an accelerated universe and thus a non-vanishing effective gravitational constant $\Lambda_\text{eff}$ \citep{Ade2016}:
\begin{equation}
\frac{\Lambda_\text{eff}}{8\pi\,G_\text{N}}
\approx 2.6 \times {10}^{-47} \, \text{GeV}^4
\approx 26  \, \text{meV}^4
\approx 6.0 \times {10}^{-27} \, \frac{\text{kg}}{\text{m}^3}
\end{equation}
whose energy scale is on the lower side of the standard model particles and similar to the estimated neutrino mass $m_\nu$. Being 55 orders of magnitude smaller than $\rho_\text{vac}$, the only possibility to consistently relate the theoretical and the experimental value is to choose $\Lambda_\text{b}$ with a high accuracy. This enormous fine-tuning -- the first 55 decimal digits of $\Lambda_\text{b}$ have to match exactly -- constitutes the \textit{new cosmological constant problem}. 

The cosmological constant is even more puzzling: While other kinds of matter dilute during the universe's expansion, the vacuum energy density remains constant so that it finally becomes the dominant component. Today, $\Lambda_\text{eff}$ is already the dominant contribution and accounts for about 70\% of the total energy density. However, it is almost of the same order as non-relativistic matter which contributes to nearly 30\% \citep{Ade2016}. Although these fractions may vary over a wide range of magnitudes according to the $\Lambda$CDM model, the question why we observe a ratio close to unity is coined the \textit{coincidence problem}. It could be a hint towards the idea of the so-called backreaction which hypothesizes that structure formation in the universe, i.e.~matter accumulation into galaxies and galaxy clusters, could cause the cosmic acceleration.

In addition, the authors of \citep{AfshordiN2015} recently pointed out the possibility of a further problem stemming from fluctuations in the vacuum energy density. In contrast to the old cosmological constant problem, here the correlations of the energy-momentum tensor at different spacetime events lead to an additional non-vanishing energy contribution in the vacuum. This problem is dubbed the \textit{cosmological non-constant problem}.

\section{Gravitational aether scenario}
\label{sec:CCP_GravitationalAetherScenario}
There are several attempts to address the cosmological constant problems in literature, especially by modifying the Einstein field equations either on the matter side (e.g.~by introducing a new scalar field in quintessence and $k$-essence models to explain the accelerated expansion) or on the geometric side (e.g.~by changing the gravitational interaction in $f\!\left(R\right)$ gravity and scalar-tensor theories as well as other mechanisms \citep{ArkaniHamedDDG2002,DvaliHK2007}). A detailed treatment is given e.g.~in the book \citep{AmendolaT2010}. A particular elegant idea to phenomenologically solve the old cosmological constant problem has been suggested in \citep{Afshordi2008}: the gravitational aether scenario (GA) which belongs to the matter-modifying theories. 

The basic idea is to decouple the vacuum energy from the universe's geometry. For this purpose an additional term is inserted only on the matter side of the Einstein field equations which exactly cancels the vacuum contribution. Because of the special form of the vacuum energy-momentum tensor \eqref{eq:EpTVac}, the correcting term up to classical contributions can be isolated by using the trace of the ordinary energy momentum tensor $T_{\mu\nu}$, $T_\alpha{}^\alpha = 4 \rho_\text{vac} +T_{\text{class}\,\alpha}{}^\alpha$. Therefore $\frac{1}{4}\,T_\alpha{}^\alpha\,g_{\mu\nu}$ is subtracted.

However, on the one hand we require that there is neither a source nor a drain for energy or momentum, they are supposed to be conserved. On the other hand we want the Einstein tensor to be compatible with the Bianchi identities, i.e.~we want a torsion-free spacetime. These demands are expressed mathematically as the vanishing covariant divergences of $T_{\mu\nu}$ and $G_{\mu\nu}$. Both of them cannot be met at the same time until we introduce a second term on the right hand side of the Einstein field equations: $T'_{\mu\nu}$. We interpret it as the energy-momentum tensor of the so-called gravitational aether. The Einstein field equations of GA now read
\begin{equation}
G_{\mu\nu}
= 8\pi\,G' \left(T_{\mu\nu} -\frac{1}{4}\,T_\alpha{}^\alpha\,g_{\mu\nu}
  +T'_{\mu\nu}\right).
\label{eq:EFE_GAorig}
\end{equation}
This extension of GR is the most general one which complies with local covariance, linearity in the energy-momentum tensor, and elimination of vacuum energy influence. $G'$ in \eqref{eq:EFE_GAorig} is treated as the fundamental gravitational constant in contrast to $G_\text{N}$ in the ordinary Einstein field equations \eqref{eq:EFE_wGwoL}. 

The equation which causes the need to introduce the gravitational aether determines at the same time how it couples to conventional matter,
\begin{equation}
\nabla_\nu\,T'_\mu{}^\nu
= \frac{1}{4}\,\nabla_\mu\,T_\alpha{}^\alpha{}.
\end{equation}
Thus the gravitational aether's energy and momentum are not conserved, but sourced by ordinary (non-relativistic) matter. However, this equation leaves open which kind of matter the gravitational aether consists of. Afshordi chose the gravitational aether to be a perfect fluid just like the usually considered types of cosmological matter. It possesses a pressure $p'$ and an energy density $p'/\omega'$ where $\omega'$ is called the equation of state parameter. 

This modified theory of gravity alters the history of the universe. Conclusions on the duration of the radiation-dominated era from big bang nucleosynthesis and light elements' abundances, cosmic microwave background signatures, and implications on the expansion behavior by redshift observations favor large values of $\omega'$. 
From now on we assume $\omega' \to \infty$. This case describes an incompressible fluid of constant (vanishing) energy density, but non-zero pressure $p'$. This limit is named the cuscuton fluid because it is the thermodynamic analog of the cuscuton field \citep{AfshordiCG2007,AfshordiCDG2007,Afshordi2008} which stands for a family of $k$-essence scalar fields. The cuscuton field exhibits remarkable properties:
First, even though in general it has a non-vanishing momentum field it is always possible to find a coordinate system in which the volume element of the phase space vanishes. This means that there is no local dynamics and the local entropy is zero. Thus the introduction of a cuscuton field into a theory does not alter the number of the system's degrees of freedom. This aesthetic feature makes it particularly appealing since it is common sense that a theory can solve any problem if it just has enough parameters. In this way theories should be designed to contain as few degrees of freedom as possible (cf.~Ockham's razor). 
Second, as a consequence of the lacking local dynamics, causality is not violated in spite of the infinite speed of sound.
Third, in the absence of own dynamics the cuscuton field follows the fields to which it couples. This tracking behavior is the reason for its name: Dodder, scientifically cuscuta, is a parasitic plant.

The parasitic property of the gravitational aether in the cuscuton limit simplifies the modified Einstein equations \eqref{eq:EFE_GAorig}. They can be expressed in the original form \eqref{eq:EFE_wGwoL} which only contains the ordinary matter if we replace the gravitational constant $G_\text{N}$ by an effective gravitational constant $G_\text{eff}$. For the moment, we assume that there is only one type of matter with equation of state parameter $w$ present and we disregard the bare gravitational constant $\Lambda_\text{b}$. $G_\text{eff}$ solves the old cosmological constant problem since it already decouples the vacuum energy density. It is defined by
\begin{equation}
G_\text{eff}\!\left(w\right)=\frac{3}{4}\,\left(1+w\right)\,G'.
\label{eq:G_eff}
\end{equation}
As a consequence, the effective gravitational constant changes in time accoding to the dominant type of matter. During the radiation-dominated epoch we find $G_\text{R} \equiv G_\text{eff}\!\left(w=1/3\right) = G'$, while the effective gravitational constant in the matter-dominated epoch reads $G_\text{N} \equiv G_\text{eff}\!\left(w=0\right) = 3G'/4$ and is identified with the Newtonian gravitational constant $G_\text{N}$ measured today. 

Before we go on to discuss the gravitational aether scenario we ask ourselves why we call this type of matter aether. Aether originally denoted a fixed medium penetrating the whole universe which allowed light to propagate like acoustic waves spread e.g.~in water. Therefore it predicted that observers in relative motion to this so-called luminiferous aether measured a different speed of light -- in conflict with experiments like that of Michelson and Morley. The theory of special relativity states that there is no omnipresent fluid to carry light waves. Yet the aether concept returned: According to \citep{JacobsonM2004} it distinguishes a preferred reference frame, in which it is at rest. In other words, aether refers to a dynamical background field which violates local Lorentz covariance. 
General covariance is nevertheless not affected because the system's dynamics is independent of the coordinates and the aether is a dynamical field, so that the theory is in line with general relativity \citep{Mattingly2005}.

Although the gravitational aether concept was designed to solve the old cosmological constant problem, it is also capable of addressing the new cosmological constant problem \citep{PrescodWeinsteinAB2009}. Let's assume a scenario comprising a spherically symmetric black hole in which all conventional matter is confined, while gravitational aether distributes over the whole spacetime. Then this gravitational aether enhanced black hole (GABH) solution of the modified Einstein equations \eqref{eq:EFE_GAorig} is similar to the Schwarzschild black hole being the standard one in the ordinary theory. 

In contrast however, the new black hole description shows a diverging behavior in the time component of the metric in the vicinity of the horizon and at infinite distance, corresponding to high and low energies, UV and IR, respectively. This deviation is caused by the pressure $p\!\left(r\right)$ of the gravitational aether, whose value scales with the integration constant $p_0$. For $p_0=0$, the solution reproduces the well known Schwarzschild case. Further investigation reveals that the event horizon lies at a larger radius than the Schwarzschild radius $r_\text{S}=2G_\text{N}\,m$ where $m$ denotes its mass. Furthermore, we encounter the curvature singularity not in the black hole's center, but at the event horizon. This geometry reminds of fuzzballs, a black hole concept inspired by string theory in which matter in form of strings extends to the event horizon and no curvature singularity occurs in its center \citep{Mathur2005,SkenderisT2008}.

The gravitational aether black hole spacetime is capable of mimicking the current cosmological acceleration. For large distances the time component of the metric resembles a de Sitter universe, i.e.~a flat spacetime which undergoes an exponential acceleration as in the $\Lambda$-dominated universe. Comparing the weak field limits, we obtain a relation between the dark energy density $\rho_\Lambda$ and $p_0$, $p_0 = -\frac{2}{3}\,\rho_\Lambda$. Since the integration constant $p_0$ relates this far distance behavior directly with the behavior close to the event horizon, it can be used to adjust the black hole's properties. It is generally assumed that our current descriptions break down at last when quantities reach the Planck scale. For these regimes, quantum gravity is needed which e.g.~could give rise to fuzzballs. Thus the authors of \citep{PrescodWeinsteinAB2009} conjecture that the highest possible temperature $T_\text{max}$ in a local rest frame is of the order of the Planck temperature $T_\text{P}$, $T_\text{max} = \theta_\text{P}\,T_\text{P}$, where $\theta_\text{P}$ is the so-called Trans-Planckian parameter. Comparing this maximum temperature with the Hawking temperature of black hole evaporation relates $p_0$ with the black hole's mass $m$. Finally, we find
\begin{equation}
m \simeq 85\,\theta_\text{P}^{-1/3}\,M_\odot
\end{equation}
where $M_\odot$ denotes the solar mass. Thus if we lived in a gravitational aether universe with a black hole of the type discussed above, we would perceive an accelerated expansion away from the black hole. If in addition the black hole had a mass $m$ of around $85$ times the mass of the sun, this acceleration would match with cosmological observations and thus would solve the new cosmological constant problem. This description can be extended to include multiple black holes or rotating ones.

The distance between the Schwarzschild and the GABH event horizon is of the order of a Planck length $l_\text{P} \approx 1.6 \times 10^{-35}\,\text{m}$. It was hypothesized in \citep{AbediDA2016,AbediDA2017} that a gravitational wave signal could be reflected several times by Planckian structures near the horizon leading to a series of echoes. This could offer the possibility for experimental tests of the quantum nature of black holes in the future.

So far we have dealt with the old and new cosmological constant problems. Matter aggregation is a key to black hole formation which in turn leads to an accelerated expansion of the universe. This backreaction mechanism would be a natural solution of the coincidence problem, too.

However, in spite of the elegance of the GABH solution open questions remain \citep{PrescodWeinsteinAB2009}. Among them are the following: There could be substantially smaller black holes, e.g.~as a result of primordial fluctuations in the early universe. They would possess much higher Hawking temperatures leading to substantially increased pressures and cosmic accelerations being inconsistent with observations. Another issue is whether a real black hole formation can create the required gravitational aether distribution.

\section{Experimental tests}
\label{sec:CCP_ExperimentalTests}
\begin{figure}[hbt]
\centering
\includegraphics[width=0.65\linewidth]{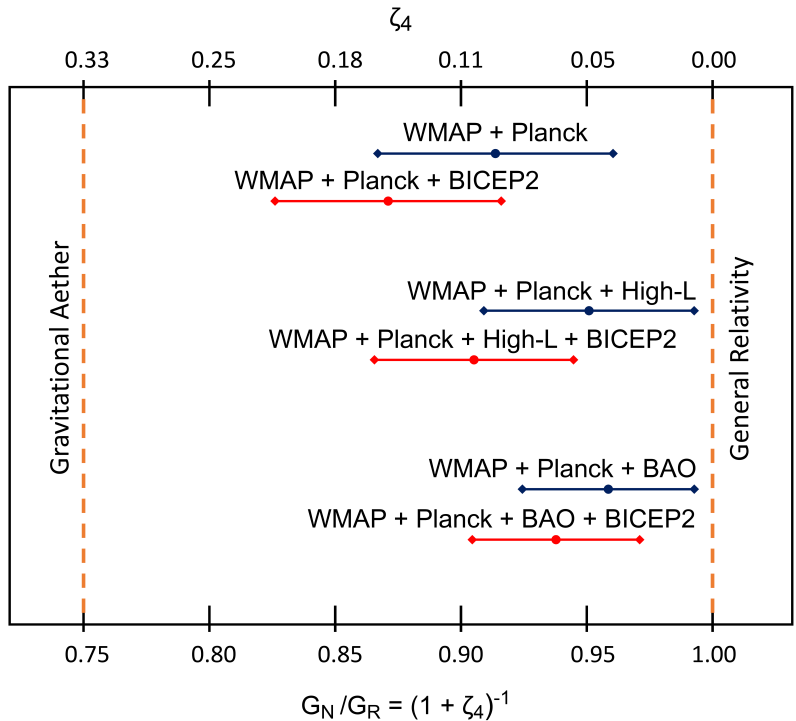}
\caption{Experimental constraints on the PPN parameter $\zeta_4$ including 1$\sigma$ error bars using baryonic acoustic oscillations (BAO), polarization of the CMB (BICEP2), high multipoles of the CMB power spectrum (High-L), CMB temperature anisotropies (Planck), and large angle polarization data of CMB (WMAP). Picture adapted from \citep{NarimaniSA2014}.}
\label{fig:Exp_zeta_4}
\end{figure}

As we have seen in \eqref{eq:G_eff}, the effective gravitational constant depends on the dominant type of matter via the equation of state parameter $w=p/\rho$ and thus differs between the radiation- and the matter-dominated era. Deviations from Einstein's general theory of relativity like this pressure dependence are cast in the so-called parametrized post Newtonian (PPN) formalism. Applying this formalism to the gravitational aether reveals that the only non-vanishing parameter is
\begin{equation}
\zeta_4
= \frac{G_\text{R}-G_\text{N}}{G_\text{N}}
= \frac{1}{3}.
\end{equation}
Thus the gravitational aether theory makes predictions which can be tested by observations of e.g.~systems with relativistic pressure or fast rotations \citep{AslanbeigiRFKA2011,NarimaniSA2014}.
In this way, experimental constraints on $\zeta_4$ can be obtained by investigating the structure of compact objects like neutron stars. However, the existing equations of state for neutron stars are yet not precise enough to allow constraining $\zeta_4$ from observed data.
Further objects to test $\zeta_4$ could be hot accretion disks of black holes and compact remnants of supernova explosions.
Focusing on the cosmological side, bounds on $\zeta_4$ can be derived from chemical element abundances due to big bang nucleosynthesis (BBN), from the cosmic microwave background radiation (CMB), or from investigations of the intergalactic matter distribution via redshifted Ly-$\alpha$ absorption lines (Ly-$\alpha$ forest).
Resulting values of $\zeta_4$ are displayed in Fig.~\ref{fig:Exp_zeta_4}. It clearly shows that the gravitational aether scenario is disfavored against Einstein's general relativity by $2.6\,\text{--}5\,\sigma$ \citep{NarimaniSA2014}.

Facing this difference with experimental results, one may look for modifications of the theory. According to \citep{NarimaniSA2014} the gravitational aether scenario could be improved in the following ways:
Above, gravitational aether has been treated as a classical thermodynamic fluid. This effective description is only valid below a certain energy scale, i.e.~it breaks down below a characteristic distance. In this picture, every particle is surrounded by a small aether halo of the order of the cut-off length $\lambda_\text{c} \sim 0.1\,\text{mm}$. This is comparable with the average baryon distance at the time of the CMB emission, which lies around $1.5\,\text{mm}$ \citep{NarimaniSA2014}.
Another approach is to keep the thermodynamic aether description, but to change its equation of state. For example one can stay with a perfect fluid, but having a non-vanishing energy density.
Furthermore, the theory could be extended to include the special role of neutrinos: Due to their small mass they behaved as radiation in the early universe and did not give rise to gravitational aether. Today, however they are non-relativistic and source gravitational aether.

\section{Summary and outlook}
\label{sec:CCP_SummaryAndOutlook}
We have seen that the cosmological constant has an interdisciplinary nature since it is connected with the fundamental concepts of general relativity, quantum field theory, and cosmology. The gravitational aether scenario is an extension of Einstein's general theory of relativity in order to decouple the vacuum energy without introducing new degrees of freedom. It offers a possible way to solve the old, but also the new cosmological constant problem and the coincidence problem. As a special characteristic, this phenomenological theory predicts testable effects in several testbeds. In its present form the gravitational aether scenario is excluded at $2.6\,\text{--}5\,\sigma$ by observations. Such a feedback can be used to develop a refined proposal. 

The observed acceleration of the universe's expansion is one of the few possible candidates for an experimentally accessible manifestation of quantum gravity. The cosmological constant problems remain unsolved which offers space for new self-consistent developments in the future. The gravitational aether scenario shows a particularly interesting proposal in the variety of ideas.

\section*{Acknowledgment}
MFW would like to gratefully thank his mentor Niayesh Afshordi for pointing his attention to the topic of gravitational aether and for many inspiring discussions about fundamental physics, especially the cosmological constant problems. MFW further wants to express his thanks to Sabine Hossenfelder, Piero Nicolini, and Ivette Fuentes as the organizers of the fifth international conference on ``Experimental Search for Quantum Gravity'' who implemented the mentoring program as a key feature.

\providecommand{\href}[2]{#2}\begingroup\raggedright\endgroup

\end{document}